\documentclass[aps,prl,twocolumn,superscriptaddress,groupaddress]{revtex4-2}

\usepackage{graphicx}  
\usepackage{dcolumn}   
\usepackage{bm}        
\usepackage{amssymb}   
\usepackage{xcolor}
\usepackage{changes}
\usepackage{comment}
\usepackage[resetlabels,labeled]{multibib}
\newcites{S}{Supplemental References}
\makeatletter

\newcommand{\Rmnum}[1]{\expandafter\@slowromancap\romannumeral #1@}
\makeatother

\makeatletter
\renewcommand{\section}{\@startsection
  {section}%
  {1}%
  {0mm}%
  {-\baselineskip}%
  {0.5\baselineskip}%
  {\normalfont\bfseries}} 
\makeatother



\begin{document}



\title{Interlayer coupling between twisted graphenes through atomically-precise barriers}

\author{Yueyang Wang}
\affiliation{Department of Physics, The University of Hong Kong, Pokfulam Road, Hong Kong, China}
\affiliation{HK Institute of Quantum Science $\&$ Technology, The University of Hong Kong, Pokfulam Road, Hong Kong, China}
\affiliation{State Key Laboratory of Optical Quantum Materials, The University of Hong Kong; Pokfulam Road, Hong Kong, China}

\author{Ben Fuller}
\affiliation{National Graphene Institute, University of Manchester; Manchester M13 9PL, UK}
\affiliation{Department of Physics and Astronomy, University of Manchester; Manchester M13 9PL, UK}

\author{Hongxia Xue}
\author{Tianyu Zhang}
\affiliation{Department of Physics, The University of Hong Kong, Pokfulam Road, Hong Kong, China}
\affiliation{HK Institute of Quantum Science $\&$ Technology, The University of Hong Kong, Pokfulam Road, Hong Kong, China}
\affiliation{State Key Laboratory of Optical Quantum Materials, The University of Hong Kong; Pokfulam Road, Hong Kong, China}

\author{Kenji Watanabe}
\affiliation{Research Center for Electronic and Optical Materials, National Institute for Materials Science, 1-1 Namiki, Tsukuba 305-0044, Japan}

\author{Takashi Taniguchi}
\affiliation{Research Center for Materials Nanoarchitectonics, National Institute for Materials Science,  1-1 Namiki, Tsukuba 305-0044, Japan}

\author{Vladimir Fal’ko}
\affiliation{National Graphene Institute, University of Manchester; Manchester M13 9PL, UK}
\affiliation{Department of Physics and Astronomy, University of Manchester; Manchester M13 9PL, UK}

\author{Dong-Keun Ki}
\email{dkki@hku.hk}
\affiliation{Department of Physics, The University of Hong Kong, Pokfulam Road, Hong Kong, China}
\affiliation{HK Institute of Quantum Science $\&$ Technology, The University of Hong Kong, Pokfulam Road, Hong Kong, China}
\affiliation{State Key Laboratory of Optical Quantum Materials, The University of Hong Kong; Pokfulam Road, Hong Kong, China}

\date{\today}

\begin{abstract}
Understanding and controlling interlayer coupling in van der Waals (vdW) materials is crucial for engineering novel electronic phenomena, including correlated states, topological phases, and superconductivity. Twisted bilayer graphene (TBG) offers a highly tunable platform to explore how interlayer orientation and separation influence quantum transport and moiré physics. Here, we introduce a method for achieving precise, atomic-scale control of interlayer coupling in TBG using dual-gated devices separated by ultrathin, thickness-tunable hBN spacers. This approach enables systematic control of interlayer interactions, demonstrating that the critical displacement field required for electron-hole bilayer behavior decreases as the layer separation increases at a fixed twist angle. Remarkably, at small twist angles, moiré-related side peaks persist even when layers are separated by tetralayers of hBN (approximately 1.3 nm), indicating robust interlayer band hybridization. By combining atomic-layer-precise control of interlayer coupling with independent twist angle tunability, our platform opens new avenues for discovering novel moiré physics governed by engineered interlayer coupling and moiré patterns.

\end{abstract}


\maketitle

%

Twistronics, a powerful platform for exploring various novel quantum phenomena, have been in investigated extensively since the study of twisted bilayer graphene (TBG)~\cite{hoppingbistritzer2011moire,tbg-insulatingcao2018correlated,tbg-superconductivitycao2018unconventional,tbg-de2021gate}. Specifically when the twist angle of two graphene layers is at the magic angle of 1.05$^\circ$, the nearly vanished Fermi velocity in the moiré-induced flat bands enhances the relative strength of electron-electron interactions, resulting in the emergence of correlated insulating states~\cite{tbg-insulatingcao2018correlated}, unconventional superconductivity~\cite{tbg-superconductivitycao2018unconventional}, and other many-body states~\cite{tbg-yankowitz2019tuning,tbg-lu2019superconductors,tbg-sharpe2019emergent,tbg-serlin2020intrinsic,tbg-xie2021fractional,tbg-he2025strongly,tbg-tanaka2025superfluid}. At larger twist angles (e.g., $\theta > 5^\circ$), the electronic bands of two graphene layers are more separated, rendering the two graphene layers electronically decoupled. In this regime, the system exhibits independent quantum transport characteristics of the two isolated monolayer graphene~\cite{decoupled-tbgkim2013breakdown,decoupled-tbgdeng2020interlayer,decoupled-tbgpiccinini2021parallel,decoupled-tbgkim2022robust,decoupled-tbgkim2023orbitally,decoupled-tbgli2024strongly}. The control of interlayer coupling in TBG has thus been achieved mostly by varying the twist angle ($\theta$), which increases the separation between the two valleys in momentum space for each graphene layer~\cite{hoppingbistritzer2011moire,tbg-superconductivitycao2018unconventional,tbg-insulatingcao2018correlated,tbg-lu2019superconductors,tbg-serlin2020intrinsic}. However, changing the twist angle intrinsically alters the moiré superlattice period, local atomic reconstruction, and strain profile, making it difficult to disentangle purely electronic coupling effects from structural modifications. Therefore, developing a platform that enables independent tuning of the interlayer coupling strength at a fixed twist angle is essential for isolating the roles of interlayer coupling and moiré patterns and systematically probing the underlying physics.

 In this context, the interlayer distance $d$ is a critical yet under-explored tuning parameter, particularly in the intermediate regime. While hydrostatic pressure has been used to reduce $d$ below $d_0 \approx0.33~\text{nm}$ (the interlayer separation in pristine TBG), this results in strong interlayer coupling and band hybridization~\cite{tbg-pressurecarr2018pressure,tbg-pressurechittari2019pressure,tbg-yankowitz2019tuning,tbg-pressurehan2025pressure,tbg-pressure-theorepadhi2019pressure,tbg-pressure-theorelin2020pressure,tbg-pressure-theorepalamara2025magic}, making independent control of each layer impossible. Conversely, increasing $d$ far beyond $d_0$ effectively decouples the layers, leading to interactions dominated by long-range Coulomb forces and negligible hybridization, as observed in Coulomb drag phenomena~\cite{CD-prlwang2024coulomb,coulombdraggorbachev2012strong,coulombdraglee2016giant,coulombdragli2016negative,coulombdragzhu2020frictional}. The intermediate regime, where interlayer coupling remains strong enough to exhibit finite moiré and interlayer band hybridization effects while the layers can be tuned separately within accessible parameters, represents a largely untapped area of study. Achieving precise control over $d$ in this range therefore would enable the realization of moiré effects together with independent layer control, opening new avenues for discovering novel electronic states and quantum phenomena in layered systems.

\begin{figure*}[t]
  \centering
  \includegraphics[width=0.98\textwidth]{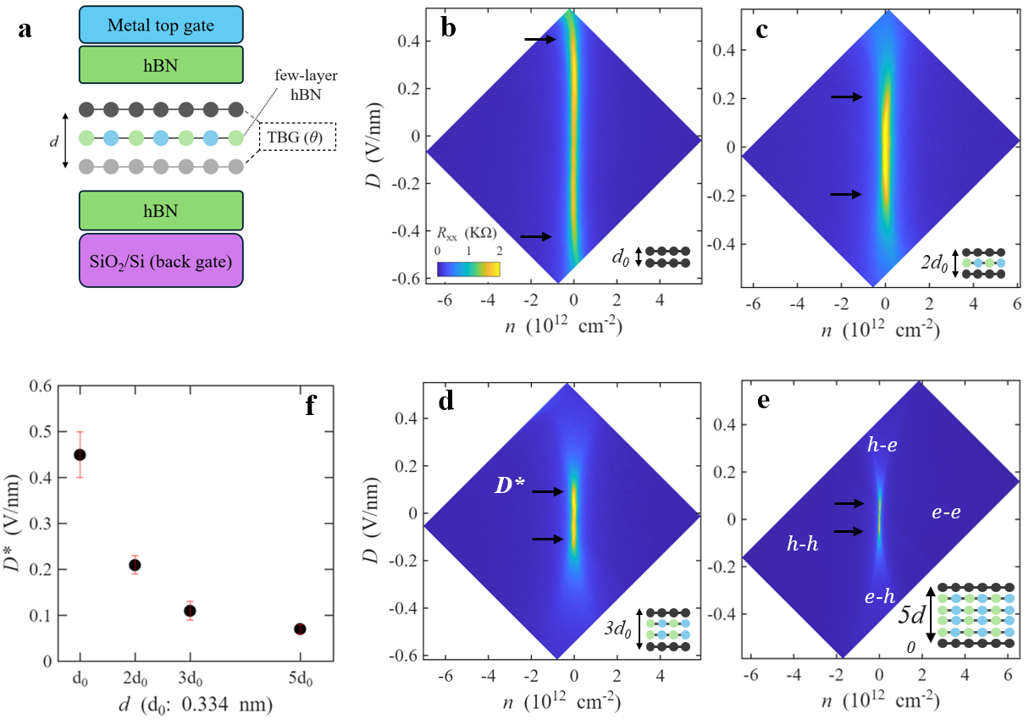}
  \caption{(a) Schematics of typical device structure. Two graphene layers are twisted with a angle $\theta$ and separated by an ultrathin hBN spacer. (b-e) The longitudinal resistance as a function of $n$ and $D$ at $B=0$ T of 0-hBN device (b), 1-hBN device (c), 2-hBN device (d), and 4-hBN device (e). Devices 0-hBN, 1-hBN, and 2-hBN are made from the same graphene flakes. The schematic inset demonstrates the interlayer distance for each device. Black arrows denote the position of the critical displacement field $D^\ast$ where the Dirac peak splits or broadens and $e-h$ bilayer is formed ($e$: electron, $h$: hole). Data in (b-d) are measured at 100 mK, and in (e) at 300 mK. (f) $D^\ast$ as a function of interlayer distance $d$, extracted from (b-e).}
  \label{fig:figure1}
\end{figure*}

\begin{figure*}[t]
  \centering
  \includegraphics[width=0.98\textwidth]{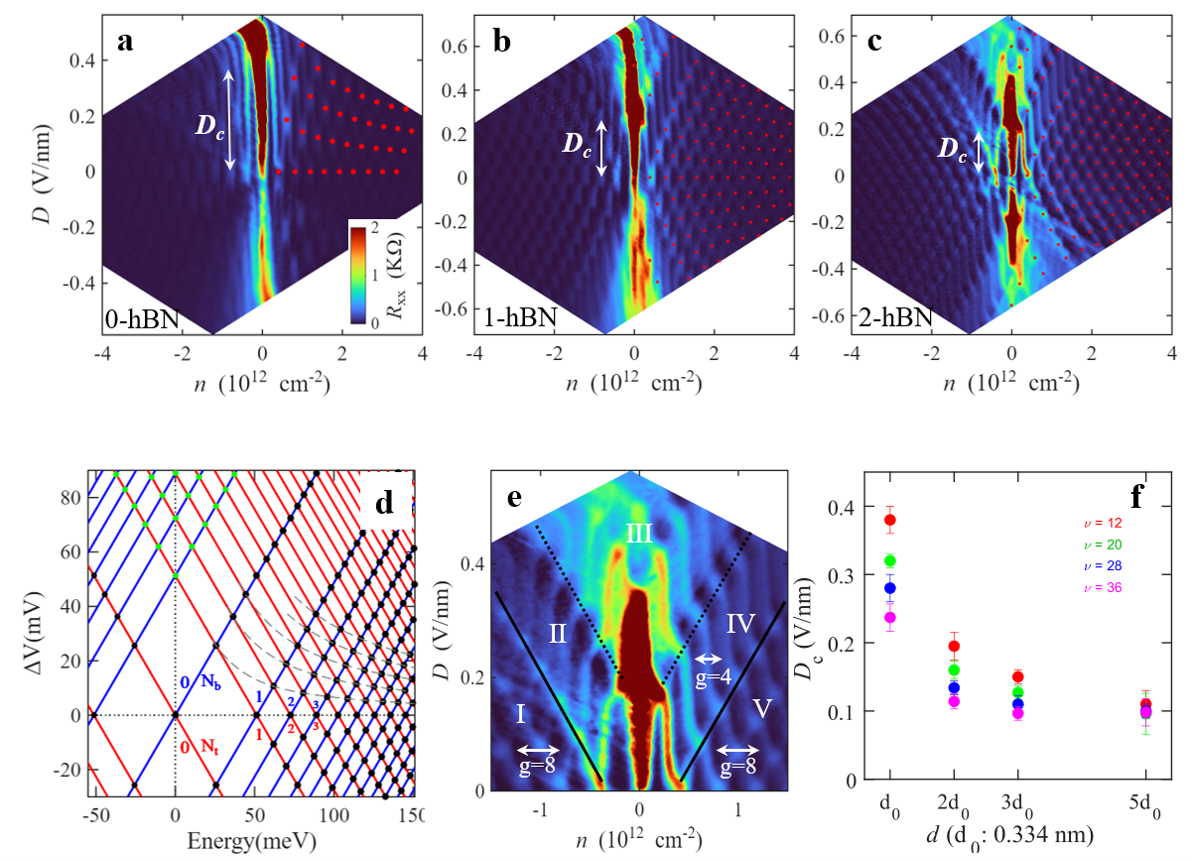}
  \caption{(a-c) Longitudinal resistance as a function of $n$ and $D$ at 2 T for 0-hBN device (a), 1-hBN device (b), and 2-hBN device (c). The red dots are calculated by the capacitance model, demonstrating a great match with experimental data. Measurement of (a) is done at 6 K while (b,c) at 100 mK. (d) The simulated crossing pattern using the equation $e\Delta V=E_{LL}(N_t)-E_{LL}(N_b)$ where $E_{LL}(N)=sgn(N)\nu_{_F}\sqrt{2e\hbar B|N|}$, $N_t$ and $N_b$ are the Landau level indices for the top and bottom layer respectively. The green dots denote crossings from electron-hole bilayers. (e) A zoom-in plot from (c) with distinct population regions of quantum oscillations. (f) The extracted displacement field between two crossings, $D_c$, as a function of $d$ at different filling factors.}
  \label{fig:figure2}
\end{figure*}

\begin{figure*}[t!]
  \centering
  \includegraphics[width=0.90\textwidth]{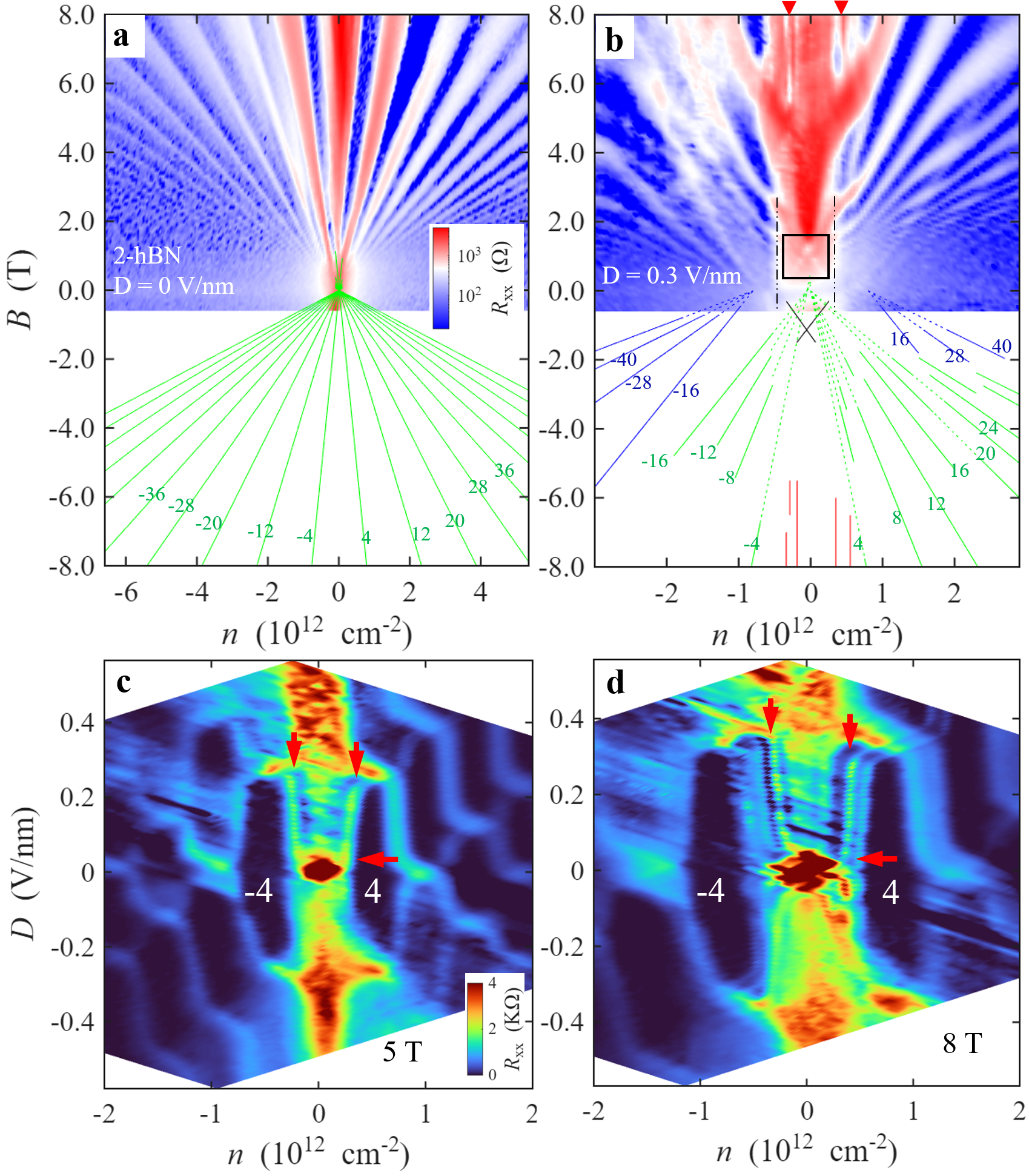}
  \caption{(a,b) Landau fan diagrams of 2-hBN device at $D=0$ V/nm (a) and $D=0.3$ V/nm (b) with schematics of filling states. Without applying $D$, the quantum oscillations exhibit 8-fold degeneracy, with filling factors $\nu=\pm4, \pm12, \pm20,...$. Under finite $D$, the degeneracy is immediately lifted, and multiple fanning points emerge. The black broken lines in (b) indicate the area of electron-hole bilayer region. The vertical features are indicated by red lines and triangles. (c,d) Longitudinal resistance of 2-hBN device as a function of $n$ and $D$ at 5 T (c) and 8 T (d). The positions of filling factors $\nu=nh/(eB)$ of $\pm4$ are indicated in white numbers. All measured at 200 mK.}
  \label{fig:figure3}
\end{figure*}

\begin{figure*}[t!]
  \centering
  \includegraphics[width=0.98\textwidth]{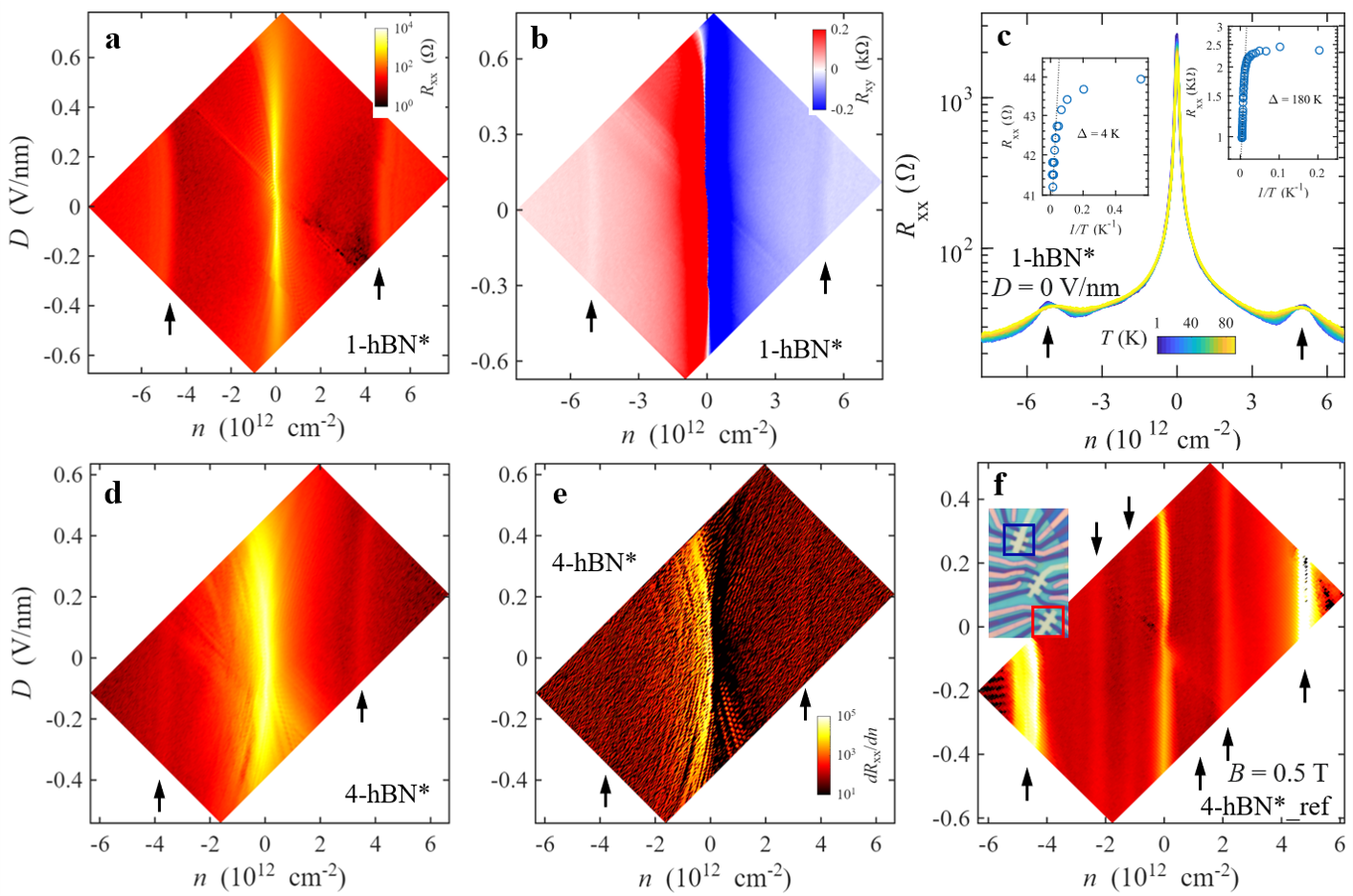}
  \caption{(a,b) $R_{\text{xx}}$ (a) and $R_{\text{xy}}$ (b) as a function of $n$ and $D$ measured at 0.3 T and 0.3 K for 1-hBN$^\ast$ device. The black arrows indicate the position of side peaks at $n_{S1}=\pm4.9\times10^{12}$ cm$^{-2}$ (1.45$^\circ$) in (a) and the corresponding Hall feature in (b). (c) Temperature-dependent $R_{\text{xx}}$ of 1-hBN$^\ast$ device measured at $D=0$ V/nm. The insets present the Arrhenius plots, where the dotted line denotes the Arrhenius fitting and $\Delta$ is the extracted gap of the side peak (left) and the main peak (right). (d) $R_{\text{xx}}$ as a function of $n$ and $D$ measured at 0.5 T and 0.3 K for 4-hBN$^\ast$ device.  The black arrows indicate the position of side peaks at $n_{S4}=\pm3.8\times10^{12}$ cm$^{-2}$, corresponding to 1.28$^\circ$. (e) Differential resistance $dR_{\text{xx}}/dn$ as a function of $n$ and $D$ for 4-hBN$^\ast$ device, exhibiting a clear signature of side peaks and the decoupled characteristics. (f) Color-plot $R_{\text{xx}}$ of the reference device relative to 4-hBN$^\ast$ device, exhibiting full, half, and quarter filling states, indicating the moiré between two graphene layers. The inset displays the optical image of the fabricated sample, where the red box represents the region of TBG separated by tetralayer hBN, while blue box denotes pristine TBG.}
  \label{fig:figure4}
\end{figure*}

In this study, we use an atomically thin and flat, highly insulating hBN flake down to the monolayer limit as an atomically precise spacer to tune the interlayer coupling in TBG without compromising device quality. By twisting two graphene layers at a fixed angle ($\theta$) separated by hBN of precisely controlled thickness~\cite{tianyuzhang2024accurate}, we achieve atomic-layer-precise control of interlayer distance $d$, while forming TBG moiré superlattices (Fig.~\ref{fig:figure1}). Dual-gate control allows independent tuning of total carrier density ($n=n_t+n_b$; $n_{t,b}$: the density of the top and bottom graphene layers) and vertical displacement field ($D$), enabling selective filling of individual layers and Landau levels at finite magnetic fields ($B$). We find that the critical displacement field, above which electron-hole bilayer behavior (i.e., $n_t\cdot n_b<0$) emerges, decreases with increasing $d$, consistent with reduced interlayer capacitance observed in Landau level crossings (Fig.~\ref{fig:figure2}). Landau fan diagrams (Fig.~\ref{fig:figure3}) further confirm electron-hole bilayer formation and exhibit vertical, non-dispersive quantum Hall features near $n=0$ at high $B$. These distinct vertical features indicate the coupling between Landau levels of electrons and holes which reside on the top and bottom graphene layers. Notably, at small twist angles, moiré-induced side peaks persist even with four-layer hBN spacers, illustrating the robust extension of moiré effects through multiple hBN layers (Fig.~\ref{fig:figure4}). These results demonstrate that high-quality, atomically thin hBN is an ideal spacer for systematic and precise control of interlayer interactions and moiré phenomena in twisted vdW systems.

The cut-and-twist transfer method is used to assemble TBG devices with or without an ultrathin hBN spacer (see Methods for details)~\cite{stackwang2013one,vdwkim2016van,cutrao2024scratching}. The typical device schematics are shown in Fig.~\ref{fig:figure1}a. The metal top gate and a highly doped silicon substrate as back gate with voltages $V_t$ and $V_b$ are used to tune carrier density as well as displacement field: $ne=(n_t+n_b)e=C_t(V_t-V_{t0})+C_b(V_b-V_{b0}), D=(n_b-n_t)e/2=[C_b(V_b-V_{b0})-C_t(V_t-V_{t0})]/2$, where $C_t (C_b)$ is the capacitance for top (back) gate, $e$ is the electric charge, and $V_{t0}$ ($V_{b0}$) is charge neutrality point of top (bottom) graphene layer. To systematically investigate the effect of $d$ on interlayer coupling, we first selected a twist angle above 2$^\circ$ to eliminate complications from moiré patterns and minimize sample-to-sample variation. To further ensure consistency, we fabricate the devices with different number of hBN spacers from the same graphene flake (Fig. S1). Figures~\ref{fig:figure1}(b-e) show the longitudinal resistance $R_{\text{xx}}$ as a function of $n$ and $D$ measured at zero magnetic field ($B=0$) for different thickness of hBN spacers; a TBG device with $d=d_0$ refer to as 0-hBN device below (Fig.~\ref{fig:figure1}(b)), TBG separated by monolayer hBN with $d=2d_0$ refer to as 1-hBN device below (Fig.~\ref{fig:figure1}(c)), TBG separated by bilayer hBN with $d=3d_0$ refer to as 2-hBN device below (Fig.~\ref{fig:figure1}(d)), and TBG separated by tetralayer hBN with $d=5d_0$ refer to as 4-hBN device below (Fig.~\ref{fig:figure1}(e)). 
 
Notably, Figs.~\ref{fig:figure1}(b-e) illustrate that along the charge neutrality line ($n=0$), $R_{\text{xx}}$ splits or broadens when $|D|$ exceeds the critical value $D^\ast$ indicated by black arrows for all four devices. This is determined by identifying the point at which $R_{\text{xx}}$ begins to drop sharply, signaling the onset of independent layer doping (Fig. S2). The observed broadening or splitting of the resistance peak effectively divides the phase diagram into four distinct regions, which correspond to different carrier population configurations in the two graphene layers. This division becomes clearer with increasing $d$, as demonstrated in Fig.~\ref{fig:figure1}(e). Apart from the $h-h$ ($e-e$) region where both layers contain holes (electrons), electron-hole ($e-h$ and $h-e$) regions emerge where electrons (holes) accumulate in one layer and holes (electrons) in the other, leading to a low-resistance state over an expanding range of $n$ as $D$ increases. This behavior originates from the decoupling of the two graphene layers due to the separation of the two graphene layers by hBN spacers. Unlike the opening of a bandgap in Bernal-stacked bilayer graphene~\cite{gapBernaloostinga2008gate,gapBernalmak2009observation,gapBernalzhang2009direct,gapBernaltaychatanapat2010electronic,gapBernalicking2022transport} or the modulation of layer polarization in conventional TBG~\cite{decoupled-tbg-screeningsanchez2012quantum,decoupled-tbgpiccinini2021parallel,layerpolaridale2023layer}, applying a displacement field $D$ above $D^\ast$ in our system shifts the Dirac points of the two layers in opposite directions, creating a moiré-coupled electron-hole bilayer.
 
As shown in Fig.~\ref{fig:figure1}(f), $D^\ast$ decreases rapidly with increasing $d$, directly reflecting the expected decay of interlayer coupling with layer separation $d$~\cite{hoppingbistritzer2011moire}. A reduced coupling strength weakens the interband hybridization between layers, lowering the energy barrier required for the displacement field $D$ to break layer symmetry. Consequently, increasing $d$ enables independent carrier doping at smaller $D^\ast$. This distinct, layer-selective carrier distribution confirms that the graphene layers are sufficiently decoupled to allow independent control at large $D$, and importantly, that this control can be tuned by adjusting the interlayer spacing $d$ with an atomic precision for a fixed twist angle to provide an ideal framework to continuously tune the moiré system into an electron-hole bilayer regime~\cite{decoupled-tbgpiccinini2021parallel,e-hbilayerrickhaus2021correlated}.

Magneto-transport studies provide further insights into the interlayer coupling effects, as summarized in Figs. 2, 3, S3 and S4. Figures~\ref{fig:figure2}(a-c) demonstrate longitudinal resistance $R_{\text{xx}}$ as a function of $n$ and $D$ at 2 T for 0-hBN, 1-hBN and 2-hBN devices, respectively. The Landau level splittings are clearly observed in $R_{\text{xx}}$ where each peak at $D=0$ V/nm splits in two for $D>0$ V/nm (or $D<0$ V/nm), reflecting the breaking of layer symmetry and the lifting of the layer degeneracy. At higher $D$, these split peaks cross with neighboring levels, producing a series of Landau level crossings between the top and bottom graphene layers. These corssing behavior is from the shift of Landau levels in graphene layers, consistent with the studies from large-angle-twisted bilayer graphene systems without hBN spacer~\cite{decoupled-tbgkim2022robust,decoupled-oddkim2021odd,decoupled-checkerboardsdong2026quantized}. This indicates that tuning interlayer distance acts same as changing twist angle but without varying moiré pattern.

These crossings can be analyzed to experimentally estimate interlayer coupling strength in the form of interlayer capacitance. To model these crossings, the heterostructures are considered as two essentially independent graphene monolayers whose Landau levels are shifted relative to each other by the potential difference $\Delta V$ induced by the application of $D$~\cite{decoupled-tbg-screeningsanchez2012quantum}. The crossings happen when $e\Delta V=E_{LL}(N_t)-E_{LL}(N_b)$ where $E_{LL}(N)=sgn(N)v_F\sqrt{2e\hbar B|N|}$, $N_t$ and $N_b$ are the Landau level indices for the top and bottom layer respectively, $\hbar$ is the reduced Planck constant and $v_F\approx1\times10^6$~m/s is the Fermi velocity. The simulated crossing pattern using the $\Delta V-E_{LL}$ equation is shown in Fig.~\ref{fig:figure2}(d). To quantitatively match the experimental crossing positions, the capacitance model is used to account for screening effects (see details in Supplementary Information section 1)~\cite{falko-slizovskiy2021out}. Excellent agreement between capacitance model and experiment, as shown by the trend in Fig.~\ref{fig:figure2}(d) compared with the trend of the red dots, (red points in Figs.~\ref{fig:figure2}(a-c) and Fig. S3(a)) is achieved with fitted capacitance values of $C_G=0.055$~F/m$^2$ (0-hBN), $0.028$~F/m$^2$ (1-hBN), and $0.020$~F/m$^2$ (2-hBN), consistent with the reduced interlayer coupling with $d$~\cite{capacitancelin2025tunneling}. In comparison, the geometric capacitance of TBG and of two graphene layers separated by mono and bilayer hBN is 0.026 F/m$^2$ (0-hBN), 0.021 F/m$^2$ (1-hBN), and 0.017 F/m$^2$ (2-hBN), respectively, all of which are smaller than the experimentally estimated values. Notably, the difference between the experimental and geometric capacitance values becomes smaller as the hBN spacer thickness increases, consistent with weakening of the interlayer coupling at larger $d$. Specifically, in TBG (i.e., the 0-hBN device), strong interlayer coupling leads to a capacitance significantly exceeding the geometric value ($0.055$~F/m$^2>0.026$~F/m$^2$), whereas increasing $d$ by inserting hBN spacer (in the 1-hBN and 2-hBN devices), the capacitance decreases and approaches to the geometric limit.

Landau level splitting also allows us to characterise layer doping in more detail than the measurements at $B=0$ (Fig.~\ref{fig:figure1}) as $R_{\text{xx}}$ varies sensitively with Landau level fillings. Figure~\ref{fig:figure2}(e) shows a detailed view near $n=0$ and $D>0$ from Fig.~\ref{fig:figure2}(c), clarifying different carrier configurations. In regions~\Rmnum{1} and~\Rmnum{5}, multiple Landau level crossings occur as $D$ varies, and the period of resistance oscillations at fixed $D$ across these crossings (indicated by white arrows), $\Delta n\approx 3.8\times10^{11}$~cm$^{-2}$, indicates a corresponding Landau level degeneracy of $g\equiv\Delta\nu\equiv\Delta n\cdot h/(eB)\approx8$ where $\nu=nh/(eB)$ is a Landau level filling factor. This degeneracy reflects the combined spin and valley degrees of freedom in both graphene layers, corresponding to the $h-h$ and $e-e$ states ($N_t\cdot N_b>0$ in Fig.~\ref{fig:figure2}(d)). In contrast, regions~\Rmnum{2} and~\Rmnum{4} show qualitatively different behavior: Landau level crossings are absent over a wide range of $D$ (equivalently, large $e\Delta V$) and the resistance oscillation period, $\Delta n\approx 1.9\times10^{11}$~cm$^{-2}$, is approximately half that in regions~\Rmnum{1} and~\Rmnum{5} (thus $g\approx 4$, as indicated by white arrows). This reduced periodicity and the large $e\Delta V$ needed to reach the next Landau level crossing are consistent with a configuration where one layer occupies its zero-th Landau level, which is energetically more isolated from other Landau levels, while the other is doped ($N_t\cdot N_b=0$). Lastly, at larger $D$ near zero $n$ (region~\Rmnum{3}), electron and hole Landau levels from top and bottom graphene layers are simultaneously filled, giving rise to electron-hole bilayer states ($N_t\cdot N_b<0$, green dots in Fig.~\ref{fig:figure2}(d)). Notably, the 0-hBN device (Fig.~\ref{fig:figure2}(a) and Fig.~S3(b)) doesn't exhibit region~\Rmnum{3} within the accesssible range of $D$ due to strong interlayer coupling.

Consequently, we can further quantify the effect of interlayer coupling on Landau level crossings by extracting the critical displacement field $D_c$, defined as the value of $D$ at which the system transitions from one Landau level crossing to the next at a fixed $n$ (i.e., at a constant $\nu$) as indicated by arrows in Figs.~\ref{fig:figure2}(a-c). Figure~\ref{fig:figure2}(f) summarizes $D_c$ as a function of $d$ for various filling factors, with values extracted from Figs.~\ref{fig:figure2}(a–c) and Fig. S3(a). For a given $d$, $D_c$ decreases with increasing $\nu$, as expected from the Landau level crossing pattern shown in Fig.~\ref{fig:figure2}(d) (see the gray dotted line). Similarly, at a fixed $\nu$, $D_c$ decreases as $d$ increases, reflecting the reduced interlayer coupling at larger $d$ and highlighting how the interlayer distance directly influences the electronic interactions between layers: weaker coupling requires a smaller potential difference $\Delta V$ (and thus a smaller $D$) to induce sufficient energy splitting between the Landau levels of the top and bottom graphene layers, allowing Landau level crossings to occur. This trend aligns well with the decrease of $D^\ast$ in $d$ shown in Fig.~\ref{fig:figure1}(f) at $B=0$.

To investigate behavior of electron-hole bilayers in region~\Rmnum{3}, we examine Landau fan diagram up to 8 T with varying displacement fields ($D$) and different numbers of hBN spacer, as summarized in Fig.~\ref{fig:figure3} and Fig. S4. First, as shown in Fig.~\ref{fig:figure3}(a) and Figs. S4(a,c), the characteristic eight-fold degeneracy of quantum oscillations is observed in all devices at $D=0$ V/nm, regardless of the hBN spacer thickness. Dominant resistance minima appear at filling factors $\nu=\pm4, \pm12, \pm20,...$, reflecting the combined degeneracy arising from spin, valley, and layer degrees of freedom when both layers are charged equally at zero $D$. In contrast, at a sufficiently large displacement field ($D=0.3$ V/nm), the influence of the hBN spacer becomes clearly evident. While 0-hBN device (Fig. S4(b)) shows a single fanning point without detectable signature of electron-hole bilayer formation, multiple fanning points and electron-hole bilayer states emerge in the Landau fan diagrams of both 2-hBN and 1-hBN devices (Figs.~\ref{fig:figure3}(b) and Fig. S4(d)) due to the applied $D$ exceeding their critical displacement field ($D^\ast\approx0.1$ V/nm for 2-hBN evice and $D^\ast\approx0.2$ V/nm for 1-hBN device, respectively). For 2-hBN device, the central fanning point at $n=0$ (green lines) arises from interlayer hybridization, while the two additional side points at $n=\pm0.8 \times 10^{12}$ cm$^{-2}$ (blue lines) correspond to the charge neutrality points of the individual graphene layers, as illustrated in Fig.~\ref{fig:figure3}(b). This coexistence highlights the decoupled yet still correlated nature of the system. Specifically, for $B<2$ T within the electron-hole region near $n=0$, we observe that $R_{\text{xx}}$ minima appear at lower $n$ as $B$ increases, marked by the black box in Fig.~\ref{fig:figure3}(b). This is opposite to the conventional Landau fan behavior, where resistance minima at fixed $\nu$ occur at larger $n$ with increasing $B$. This reversal indicates that the trajectories of Landau levels in the two layers shift in opposite directions as increasing $B$, reflecting independently populated electron and hole states. Such behavior is consistent with previous observations of correlated electron-hole states in large-angle-twisted double bilayer graphene systems~\cite{e-hbilayerrickhaus2021correlated}.

Beyond standard quantum Hall states, our high-field magnetotransport measurements uncover striking magnetic-field-independent features that signal unconventional electronic correlations in electron-hole bilayer TBG system. Specifically, Fig.~\ref{fig:figure3}(b), the fan diagram at $D=0.3$ V/nm of 2-hBN device, reveals several vertical features indicated by the red lines and triangles in the electron-hole region that corresponds to a total filling factor below 4, i.e., when both layers are occupying the zero-energy Landau levels. Unlike ordinary Landau level features, these features do not disperse in ($n, B$). Figures~\ref{fig:figure3}(c,d) further show that at fixed $B$ they appear at an intermediate range of $D$ between the two highly resistive states, which becomes wider at larger $B$, indicated by the red arrows. In this regime, electrons populate one graphene layer while holes populate the other. This leads to a situation where the zero-energy Landau level in the top layer is partially filled with electrons, while the corresponding Landau level in the bottom layer is partially filled with holes, or vice versa. As the magnetic field increases, the Coulomb interaction between these spatially separated electrons and holes becomes stronger, which can promote the formation of coupled electron-hole Landau level states. While further investigation is necessary to determine the precise origin of the observed vertical features, these results clearly demonstrate that tuning the interlayer coupling in TBG enables access to new, emergent many-body Landau level states.

After experimentally realizing atomic-level tuning of interlayer coupling in TBG, we fabricated TBG devices with hBN spacers at smaller twist angles (target below $2^\circ$), bringing the graphene bands closer in momentum space and allowing moiré effects to emerge in an experimentally accessible range of $n$. In device with monolayer hBN spacer, 1-hBN$^\ast$, we found a set of additional resistance peaks at finite $n$ on both electron and hole sides, as marked by black arrows in Figs.~\ref{fig:figure4}(a-c) along with Landau level crossings and even-odd integer quantum Hall features originating from interlayer coupling (Fig. S5). For 1-hBN$^\ast$ device, we attribute the observed side peaks ($n_{S1}=\pm4.9\times10^{12}$ cm$^{-2}$ corresponding to 1.45$^\circ$ twist) to moiré effects between the two graphene layers because a similar twist angle ($\theta \approx 1.25^\circ$) found in a reference device (1-hBN$^\ast$\_ref, Fig. S6), fabricated from the same graphene flake. The possibility of accidental graphene-hBN alignment is ruled out, supported by three key observations: (1) $R_{\text{xy}}$ exhibits only weak variation at $n_{S1}$, consistent with a weak moiré potential arising from graphene–graphene alignment across an hBN spacer, rather than strong direct graphene–hBN alignment; (2) the side peaks are symmetrically positioned and show nearly identical amplitudes, unlike the asymmetric features seen in graphene–hBN moiré systems~\cite{sdpemergenceyankowitz2012emergence,sdpwang2015evidence,sdpli2018charge}; and (3) the thermal activation gap ($\Delta_{\text{mp}} = 180\text{ K}$, right inset of Fig.~ \ref{fig:figure4}(c)) at $n=0$ is much smaller than expected for graphene-hBN moiré~\cite{angle-gaphunt2013massive}. 

Interestingly, side peaks persist in the 4-hBN$^\ast$ device with a 4-layer hBN spacer. As shown in Fig.~\ref{fig:figure4}(d), the relatively weak satellite resistance peaks appear at $n_{S4}=\pm3.8\times10^{12}$ cm$^{-2}$ (corresponding to a twist angle of 1.28$^\circ$), accompanied by evidence of enhanced electronic decoupling between the graphene layers as revealed by multiple features dispersing in $n$ and $D$ in the $dR/dn$ map (Fig.~\ref{fig:figure4}(e)). A reference sample without an hBN spacer (4-hBN$^\ast$\_ref), fabricated from the same graphene flakes, displays closely matching full-filling densities ($n_{S4ref}=\pm4.5\times10^{12}$ cm$^{-2}$) and twist angle, $\theta \approx 1.39^\circ$ (Fig.~\ref{fig:figure4}(f)), suggesting the observed side peaks could originate from the moiré superlattice potential between the graphene layers. Alternatively, these side peaks may also result from Dirac band hybridization, which creates anticrossing bands at finite energy for a given twist angle and leads to side peaks at $n$ depending on this angle, $n=2k^2_F/\pi=\Delta K^2/2\pi$, where $\Delta K^2=|k_{F1}+k_{F2}|^2=(|K|\theta)^2=8\pi^2/3\sqrt{3}A_m$, (see details in Supplementary Information section 2). Using $n_{S4}$ from Fig.~\ref{fig:figure4}(d), we estimate a twist angle, 1.6$^\circ$, slightly larger than that of the reference device. While band hybridization, arising from electron hopping between layers, is generally less sensitive to interlayer spacing than moiré superlattice effects from electron orbital coupling, the similar twist angles estimated for both the 4-hBN$^\ast$ and reference devices (assuming a moiré origin) suggest that the observed side peaks may result from a combined influence of residual band hybridization and long-range moiré potential. Overall, this platform enables tunable electronic decoupling and spatial potential engineering, providing new opportunities for exploring correlated and topological phases in van der Waals heterostructures.

In conclusion, we demonstrate atomic-scale control of interlayer distance and associated coupling in twisted bilayer graphene by fabricating high-quality encapsulated heterostructures in which two graphene layers, twisted by a small angle, are separated by an ultrathin hBN spacer of an atomically precise thickness. At zero magnetic field, the critical displacement field $D^\ast$ above which electron–hole bilayer behavior emerges decreases systematically with increasing interlayer distance $d$, directly reflecting the weakening of interlayer coupling. Landau level crossing features at finite magnetic field, together with the evolution of the Landau fan diagrams with $d$ and $D$, corroborate this trend, confirming that a twisted electron-hole bilayer forms within experimentally accessible ranges of $D$ and $n$ that depends sensitively on $d$. Within the electron-hole bilayer regime, we identify vertical quantum-Hall features that cannot be accounted for by a simple single-particle Landau level spectrum, pointing to emergent many-body or interaction-driven physics. At smaller twist angles, we observe moiré-related satellite peaks in devices with hBN spacers up to four layers thick, demonstrating that the moiré potential and the associated interlayer band hybridization can propagate coherently through several layers of hBN. Our thickness-tunable approach thus enables a systematic exploration of interlayer coupling physics and, combined with moiré engineering, offers new avenues for tailoring correlated states in two-dimensional heterostructures.

\begin{acknowledgments}
The work is financially supported by the National Key R~\&~D Program of China (2020YFA0309600) and by the University Grants Committee/Research Grant Council of Hong Kong SAR under schemes of Area of Excellence (AoE/P-701/20), CRF (C7037-22G), and GRF (17309722, 17301424, and 17300725). K.W. and T.T. acknowledge support from the JSPS KAKENHI (grant nos. 21H05233 and 23H02052), the CREST (JPMJCR15F3), JST, and the World Premier International Research Center Initiative (WPI), MEXT, Japan.
\end{acknowledgments}

\bibliography{LC}

\appendix

\clearpage
\setlength{\belowcaptionskip}{-0.1cm}
\onecolumngrid
\begin{center}
\section{\MakeTextUppercase{Supplementary Materials for Interlayer coupling between twisted graphenes through atomically-precise barriers}}
\end{center}
\vspace{5\baselineskip}

\renewcommand{\thefigure}{S\arabic{figure}}
\renewcommand{\thepage}{S\arabic{page}}
\renewcommand{\theequation}{S\arabic{equation}}
\renewcommand{\bibnumfmt}[1]{[S#1]}
\renewcommand{\citenumfont}[1]{S#1}

\setcounter{equation}{0}
\setcounter{figure}{0}
\setcounter{page}{1}
\setcounter{enumiv}{0} 
\newpage
\section{1. Device fabrication}
\begin{figure}[t]
  \centering
  \includegraphics[width=0.95\textwidth]{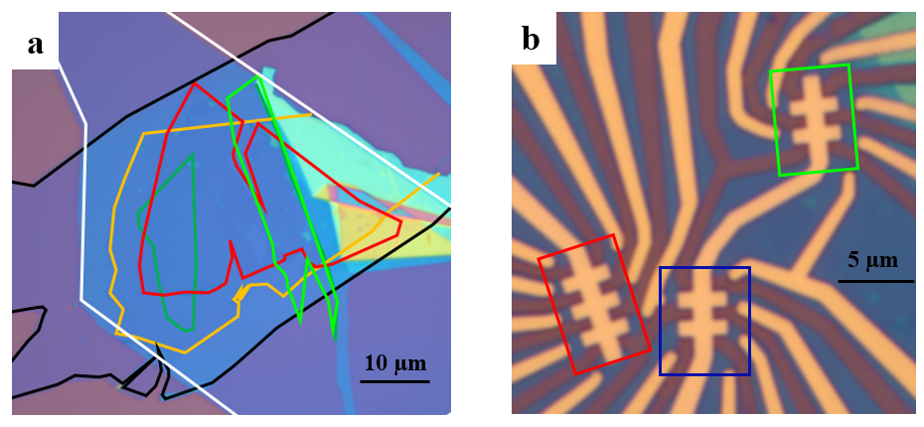}
  \caption{(a) Optical image of a typical heterostructure, overlaid with stacking configuration. The white, red, green, orange, and black sketches denote the edges of the top hBN, top graphene, middle hBN, bottom graphene, and bottom hBN, respectively. The light (dark) green marks monolayer (bilayer) hBN. (b) Optical image of the final device geometry with three different areas: pristine TBG (0-hBN device, blue box), TBG separated by monolayer hBN (1-hBN device, red box), and TBG separated by bilayer hBN (2-hBN device, green box).}
  \label{fig:figureS1}
\end{figure}

Graphene and hexagonal boron nitride (hBN) flakes were mechanically exfoliated from bulk crystals onto silicon wafers coated with a 285 nm SiO$_2$ layer using the Scotch-tape method. The hBN crystals were supplied by K. Watanabe and T. Taniguchi, while the graphite originated from NGS. Monolayer graphene flakes were first identified under an optical microscope. Accurate determination of ultrathin hBN layer numbers was achieved by combining optical contrast with second-harmonic generation (SHG) spectroscopy. The top and bottom hBN flakes were selected to be approximately 30 nm thick, as estimated optically. Once suitable flakes were identified, the heterostructure was assembled using a PDMS micro-dome coated with a polycarbonate (PC) film at 100 ℃. During pickup, we put two substrates onto the transfer stage, with one substrate having two halves of cut graphene and the other ultrathin hBN spacer, so that we record the position and angle of picking up the first half of the graphene flake and then move back to the recorded information to accurately control the twist angle between two graphene flakes with hBN spacer in the middle. All the electrical contacts are connected to the two graphene layers together. The mismatch alignment is carried out between two graphene layers with a target of a small angle ($\theta<3^\circ$), while all other hBN flakes are intentionally rotated for 15$^\circ$ with regard to both graphene layers to avoid unwanted superlattice potential between graphene and hBN flakes. Cr/Au (5 nm/50 nm) electrodes were then deposited on top of the stack using standard electron-beam lithography, electron-beam evaporation, and lift-off processes to serve as the top gate. Subsequently, one-dimensional edge contacts were made to both the top and bottom graphene layers. Finally, the encapsulated heterostructure was etched into a Hall-bar geometry. Figure~\ref{fig:figureS1} shows the typical heterostructure assembly and the resulting device geometry.

Transport measurements were performed in a $^3$He cryostat (300 mK) and in a dilution system (100 mK) both integrated with superconducting magnets. Four-terminal resistance measurements were carried out using a low-frequency lock-in technique with a frequency of 17.7 Hz. The d.c. gate voltages were output by Keithley 2400 and DC205 voltage source meters.

\section{2. Determination of $D^\ast$}
\begin{figure}[t]
  \centering
  \includegraphics[width=0.5\textwidth]{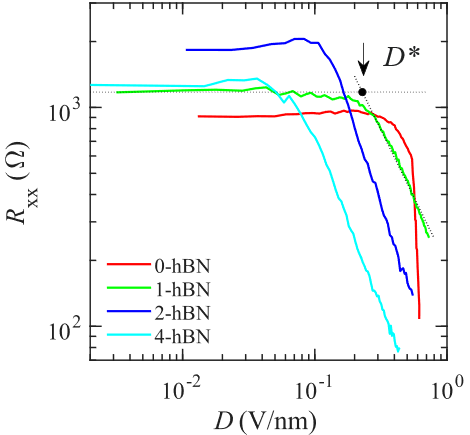}
  \caption{The log-scale longitudinal resistance as a function of $D$ at $n=0$ of the 0-hBN device, 1-hBN device, 2-hBN device, and 4-hBN device to extract the critical displacement field $D^\ast$ where the Dirac peak splits or broadens.}
  \label{fig:figureS2}
\end{figure}

$D^\ast$ is determined by the log-scale longitudinal resistance as a function of $D$ at $n=0$ for the 0-hBN device, 1-hBN device, 2-hBN device, and 4-hBN device, as demonstrated in Fig.~\ref{fig:figureS2}.

\section{3. More information for Landau level crossings}
\begin{figure}[t]
  \centering
  \includegraphics[width=0.95\textwidth]{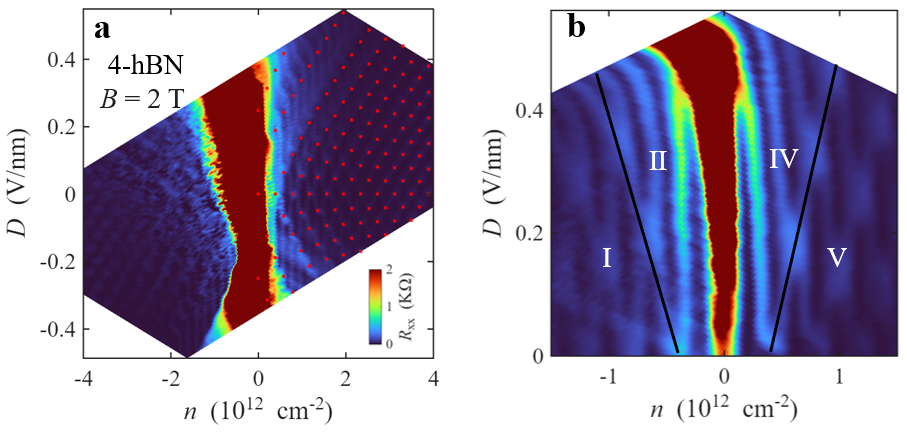}
  \caption{(a) $R_{\text{xx}}$ as a function of $n$ and $D$ at 2 T for the 4-hBN device with calculated crossings indicated by red dots. (b) A zoom-in plot of Landau level crossings in 0-hBN device at 2 T, demonstrating regions~\Rmnum{1},~\Rmnum{2},~\Rmnum{4} and~\Rmnum{5} without~\Rmnum{3}.}
  \label{fig:figureS3}
\end{figure}

$R_{\text{xx}}$ as a function of $n$ and $D$ for the 4-hBN device with calculated crossings and a zoom-in plot of Landau level crossings in 0-hBN device at 2 T are shown in Fig.~\ref{fig:figureS3}.

In this study, the graphenes are rotated relative to each other by some angle. The hBN layers are misaligned to the graphenes. Since hBN is an insulator, the only effect of the hBN layers is to decrease the tunnelling amplitude between graphenes and to increase the electrostatic potential drop between layers since the displacement field acts over a longer distance with modified permittivity. When an external displacement field is applied, the two layers to accumulate opposite charges. These charges also screen the field according to the polarisability of the graphene orbitals. The potential on each layer is found by calculating the potential difference between layers given by
\begin{equation}
e\Delta V= \frac{eD}{\varepsilon_0}(\frac{d_z}{\varepsilon_z}+N\frac{d_{hBN}}{\varepsilon_{hBN}})+\frac{e^2}{\varepsilon_0}(\frac{1+\varepsilon_z}{2})\frac{d_z}{\varepsilon_z}\frac{(n_t-n_b)}{2}
\end{equation}
where $d_z$ and $d_{hBN}$ are the thicknesses of graphene and hBN, $\varepsilon_z$ and $\varepsilon_{hBN}$ are dielectric permitivities of graphene and hBN, $N (>0)$ is the number of hBN layers.

The two graphene layers are almost independent of each other, or only very weakly coupled. This is due to two reasons: the hBN increases the distance to hop between top and bottom, and the twist angle $\theta$ between graphene layers further decouples them. Therefore, the dispersion is that of a Dirac cone in each layer separated in momentum space by $\Delta K=\theta K$ and with 2 valleys. The weak coupling is not enough to support moiré minibands. So there is an analytic relationship between electron
density and Fermi wavevector $k_F=\sqrt{\pi |n_i|}\text{sign}(n_i)$ and conical dispersion relation $\mu-E_i=\hbar vk_{Fi}$ where $i$ denotes one of the two graphene layers and $\mu$ is the chemical potential which is constant across the device. For the case $N = 0$ without any hBN, these assumptions are invalid because the layers are directly contacting and strongly coupled, so hybridisation significantly modifies the band structure.

In the presence of a magnetic field, the density of states separates into discrete Landau levels. The occupation of each level is
\begin{equation}
    \rho_{LL}=\frac{1}{2\pi \lambda^2_B}
\end{equation}
with magnetic length $\lambda_B=\sqrt{\frac{\hbar}{eB}}$ and there is an additional factor of 4 for spin and valley degeneracy. When the chemical potential coincides with a Landau level in each layer, there is resonant tunnelling between the layers and a peak in conductivity. This occurs at discrete points of total density and displacement field. These points are calculated by setting the density in each layer to be an integer multiple of the LL occupation and degeneracy and then reconstructing the displacement field necessary for such a condition, accounting for the screening by the determined electron densities by rearranging equation S1.

Without hBN spacer, the relation between potential difference $\Delta V$ and $D$ at a crossing is simply associated through $\Delta V=(D-e\Delta n/2)/C_G$, where $\Delta n=(N_t-N_b)4eB/h$, $C_G$ is the interlayer capacitance per unit area.

\section{4. Landau fans for the 0-hBN and 1-hBN devices}
\begin{figure}[t]
  \centering
  \includegraphics[width=0.95\textwidth]{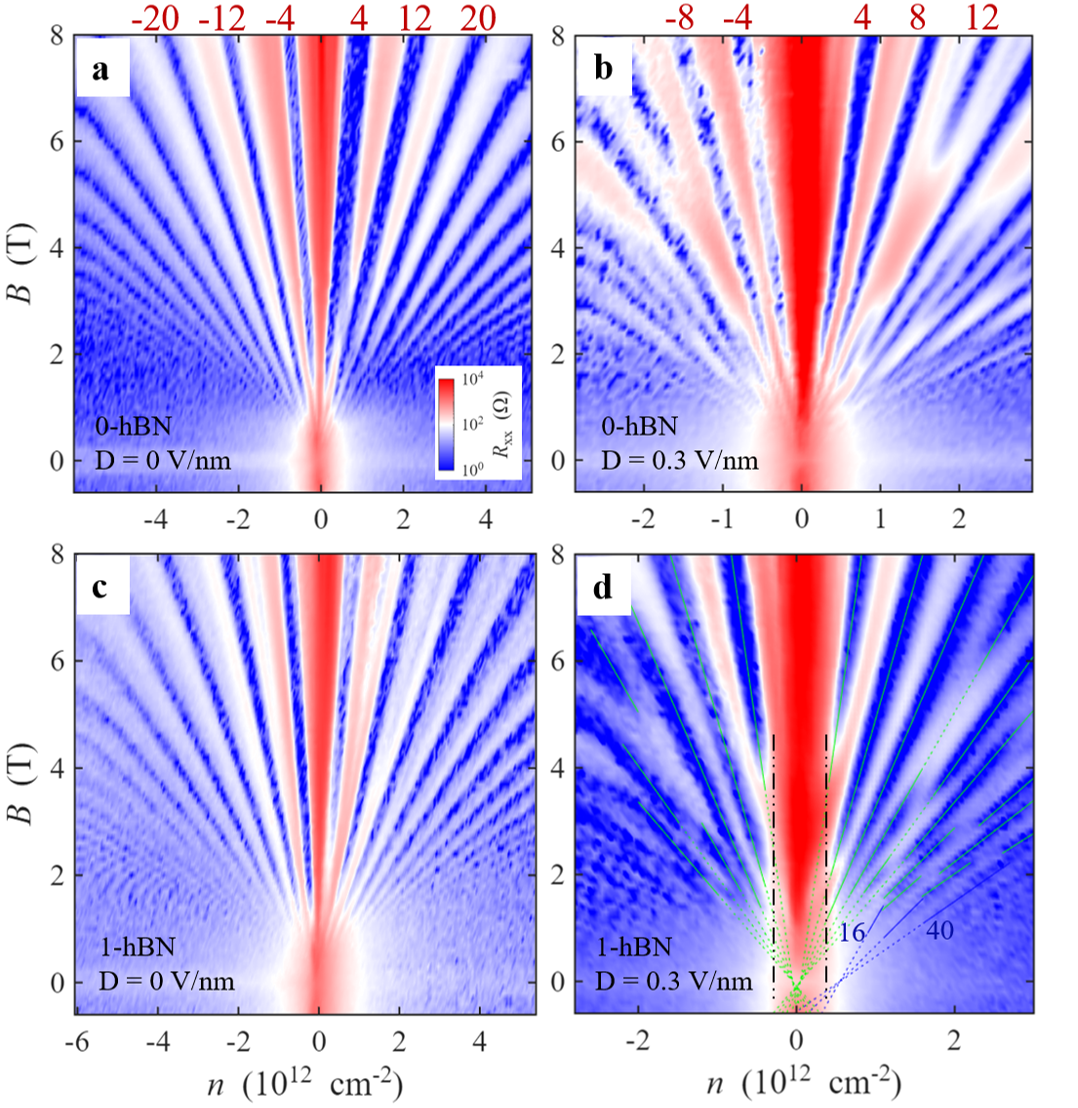}
  \caption{(a,b) Landau fans of 0-hBN device at $D = 0$ V/nm (a) showing 8-fold degeneracy (spin, valley, layer) states and at $D = 0.3$ V/nm (b) exhibiting 4-fold degeneracy with lifted layer degree of freedom. ((c,d) Landau fans of 1-hBN device at $D = 0$ V/nm (c) demonstrating 8-fold degeneracy and at $D = 0.3$ V/nm (d) displaying 4-fold degeneracy. In (d), mutiple fanning points are observed. The black broken lines indicate the electron-hole bilayer region.}
  \label{fig:figureS4}
\end{figure}

Figure~\ref{fig:figureS4} shows Landau fan diagrams of the 0-hBN and 1-hBN device (bare twisted bilayer graphene) at $D=0$ V/nm and $D=0.3$ V/nm. For the 0-hBN device, there is only single fanning point and no electron-hole coexistence region since the two graphene layers are highly coupled ($D^\ast\approx0.45$ V/nm). In contrast, the 1-hBN device illustrates a narrower electron-hole region than the 2-hBN device and two emanating points, with the one at $n=0$ corresponding to interlayer hybridization and the other one at $n=0.4$ cm$^{-2}$ arising from the neutrality point of one of the graphene layers. This difference compared with the 2-hBN device is attributed to a stronger coupling effect between the two graphene layers than the 2-hBN device ($D^\ast(\text{1-hBN})\approx 0.2 ~\text{V/nm}> D^\ast(\text{2-hBN})\approx 0.1~\text{V/nm}$).

\section{5. Magneto-transport for 1-hBN$^\ast$ device with moiré potential}
\begin{figure}[t]
  \centering
  \includegraphics[width=0.95\textwidth]{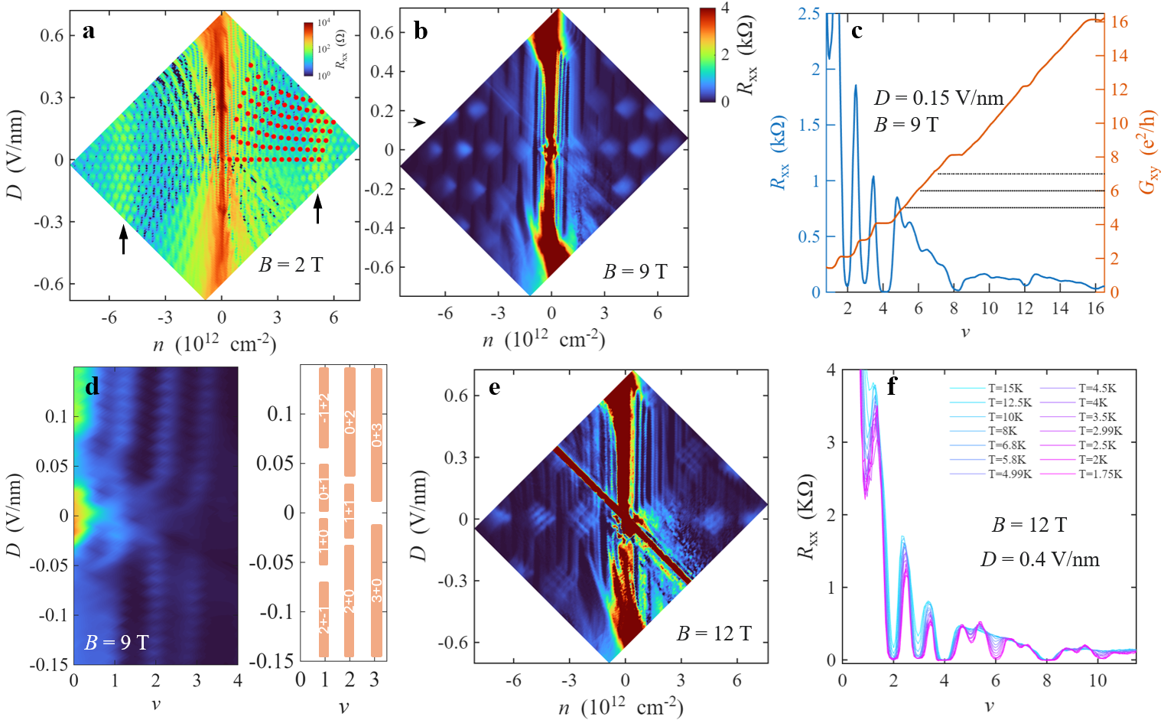}
  \caption{(a) $R_{\text{xx}}$ as a function of $n$ and $D$ at 2 T for the 1-hBN$^\ast$ device with calculated crossings indicated by red dots. (b) Color-plot of $R_{\text{xx}}$ as a function of $n$ and $D$, measured at $B = 9$ T. (c) Line cut along the black arrow in (b), $R_{\text{xx}}$ (left y-axis) at $D = 0.15$ V/nm and corresponding $G_{\text{xy}}$ (right y-axis). (d) Left: Zoom-in plot of (b) at around $|D| \leq 0.15$ V/nm and $0 \leq \nu =hn/(Be)\leq 4$. Right: Schematic of evolution with quantum Hall states in each graphene layer. The integer numbers correspond to $\nu_t$ and $\nu_b$. (e) $R_{\text{xx}}$ as a function of $n$ and $D$ measured at $B = 12$ T and $T = 0.3$ K, demonstrating $4\times 4$ checkerboards. (f) Temperature dependence measured at $D = 0.4$ V/nm and $B = 12$ T. The extracted gaps at odd-integer states are generally smaller than the gaps at even-integer quantum Hall states. For instance, $\Delta_{\nu=1} = 6$ K and $\Delta_{\nu=5} = 5$ K, while $\Delta_{\nu=1} = 18.5$ K and $\Delta_{\nu=10} = 9$ K. All features are consistent with samples without moiré.}
  \label{fig:figureS5}
\end{figure}

Quantum Hall states of the 1-hBN$^\ast$ device are consistent with the observations in devices without moiré in the main text, as displayed in Fig.~\ref{fig:figureS5}, indicating that the moiré-coupled interaction does not change the decoupling nature.

\section{6. Reference sample for 1-hBN$^\ast$ device}
\begin{figure}[t]
  \centering
  \includegraphics[width=0.95\textwidth]{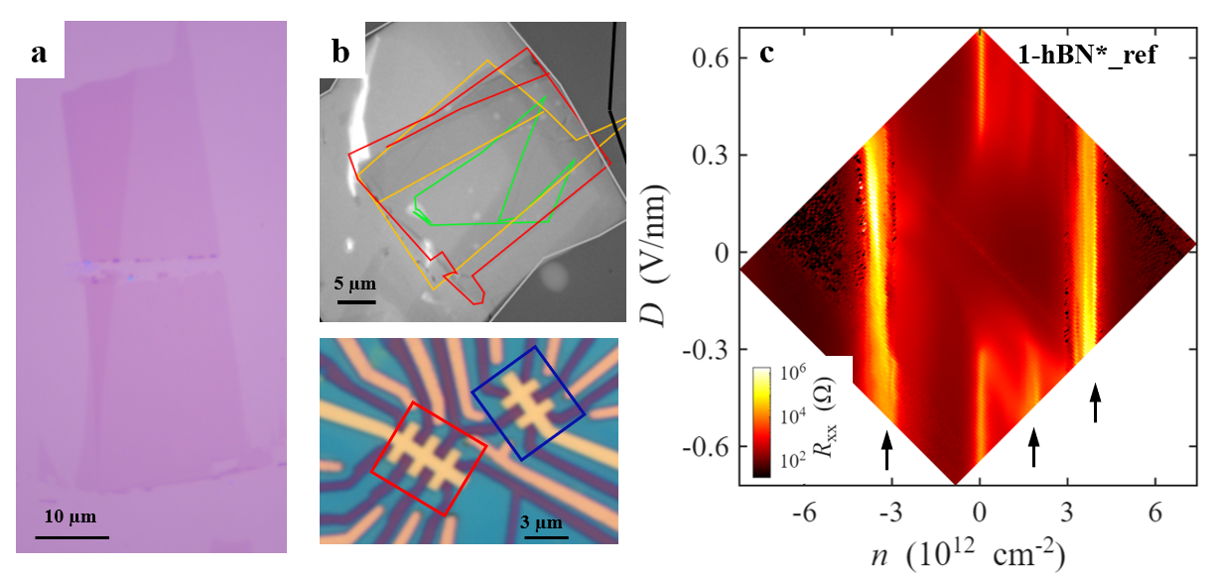}
  \caption{(a) Graphene flake used in this stacking with monolayer and bilayer regions after cutting. (b) Top: optical image (dark field) of 1-hBN$^\ast$ heterostructure, overlayed with stacking configuration. The white, red, green, orange, and black sketches denote the edges of the top hBN, top graphene, middle hBN, bottom graphene, and bottom hBN, respectively. Bottom: optical image of the fabricated sample, where the red box represents region of TBG separated by monolayer hBN while blue box denotes pristine twisted monolayer-bilayer graphene (TMBG). (c) Color-plot $R_{\text{xx}}$ of reference device TMBG relative to 1-hBN$^\ast$ device exhibiting full (at $n_{S1ref}=\pm3.65\times10^{12}$ cm$^{-2}$ corresponding to $\sim$1.25$^\circ$) and half filling states, indicating the moiré between two graphene layers.}
  \label{fig:figureS6}
\end{figure}

The reference sample with optical images and transport measurement relative to the 1-hBN$^\ast$ device is shown in Fig.~\ref{fig:figureS6}.

\section{7. Band hybridisation}
Interlayer tunnelling processes between twisted layers can be distinguished by the role of the superlattice potential. There is direct hybridisation between identical momenta, and also umklapp scattering processes which couple momenta with a boost by a moir{\'e} Bragg vector.
The latter depends on the superlattice potential while the former does not.
Therefore, the direct hybridisation between two shifted Dirac cones may survive when moir{\'e} effects are weak.
The hybridisation causes anticrossing bands around where the individual cones meet.

We can calculate the carrier density required for the Fermi level to coincide with this feature.
Consider two Dirac cones which are separated in momentum by $\Delta K = \theta K$ where $K=4\pi/3a$.
The cones touch when their Fermi wavevectors sum to this separation, $|k_{F1}+k_{F2}|=\Delta K$.
When $D=0$, the cones lie at the same energies, so the Fermi cut at the point where the cones touch is of two circles with the same radius $\Delta K/2$.
They host total density $n=2 k_F^2/\pi=\Delta K^2/2\pi$.
This can be written in units of the moir{\'e} unit cell filling, because $A_m=\frac{3\sqrt{3}}{8\pi^2}(K\theta)^2$, giving the filling factor $\nu=n A_m = \frac{4\pi}{3\sqrt{3}}\approx2.4$.
Because of the geometry that the twist angle sets the momentum displacement between cones, the filling factor at which they touch is independent of twist angle.
However, it is not an effect of the moir{\'e} superlattice and this constant filling is a coincidence.
The hybridised bands will anticross at this point and the energy gap has magnitude of the interlayer tunnelling strength.
This anticrossing occurs at positive filling, where the two cones are both electron-like, and also for negative filling, where the two cones are hole-like.
For finite displacement fields, the same condition applies but the cones have different radii, which leads to a slightly higher total density.

\newpage

\pagebreak
\newpage

\newpage

\newpage

\newpage

\newpage

\clearpage

\end{document}